# On the asymmetry of the forward and reverse martensitic transformations in shape memory alloys

**D.L. Beke, M. K. Bolgár, L.Z. Tóth, L. Daróczi**

*Department of Solid State Physics, University of Debrecen, H-4002 Debrecen, P.O, Box 2, Hungary*
*Corresponding author e-mail: dbeke@science.unideb.hu (D.L. Beke)*

**Abstract**

Differential Scanning Calorimetric, DSC, runs taken during martensitic phase transformations in shape memory alloys, often look differently during cooling and heating. Similar asymmetry is observed e.g. for the numbers of hits or the critical exponents of energy and amplitude distributions ($\varepsilon$ and $\alpha$, respectively) in acoustic emission measurements. It is illustrated that, in accordance with empirical correlations, the above asymmetry of acoustic noises can be classified into two groups: the relative changes of the exponents during cooling and heating ($\gamma_\varepsilon = (\varepsilon_h - \varepsilon_c)/\varepsilon_c$ as well as $\gamma_\alpha = (\alpha_h - \alpha_c)/\alpha_c$) are either positive or negative. For positive $\gamma$ values the number of hits and the total energy of acoustic emission are larger for cooling, and the situation is just the reverse for negative asymmetry. Our interpretation is based on the different ways of relaxation of the elastic strain energy during cooling as well as heating. It is illustrated that if the relaxed fraction of the total elastic strain energy (which would be stored without relaxations) during cooling is larger than the corresponding relaxed fraction during heating, then the asymmetry is positive. Magnetic emission noises, accompanied with martensitic phase transformations in ferromagnetic alloys, show similar asymmetry than those observed for thermal (DSC) and acoustic noises and depends on the constant external magnetic field too.

1. **Introduction**

Differential Scanning Calorimetric, DSC, runs taken during martensitic phase transformations in shape memory alloys often look differently during cooling and heating. This is especially striking if the cooling/heating rates are low enough (and the mass of the sample is also small enough) i.e. the transition is adiabatic, but still athermal [1]. For example in Fig.1 (taken from [2]), instead of having one wide envelope-like DSC curve, the DSC spectra split into a number of individual thermals spikes during cooling, while it contains only one sharp peak during heating in $Ni_2MnGa$ single crystalline samples with 10M modulated martensite structure. Similar result can be seen on Fig. 2, obtained in NiFeGaCo single crystals [3]. The presence of separate peaks makes possible the determination of the power exponents of energy, $\varepsilon$, (using that the heights of the DSC peaks are proportional to the elementary energy released or absorbed) characterizing the distribution function of the energy of individual peaks:

$$P(E) = CE^{-\varepsilon} \exp\left(-\frac{E}{E_c}\right), \tag{1}$$

where $C$ is a normalization factor, $\varepsilon$ is the critical exponent and $E_c$ is the cutoff value (see e.g. [2,3,4]). It is well-known that the martensitic transformation is an intermittent process characterized by avalanches and the validity of (1) is an indication of behaviour of driven criticality [5,6]. Beside the above thermal avalanches, avalanches of acoustic and magnetic emission signals can also be detected (see e.g. [2-11]) during martensitic transformations and the probability distribution functions of amplitude, A, size, S, and time, T, can also be characterized by an expression similar to (1).

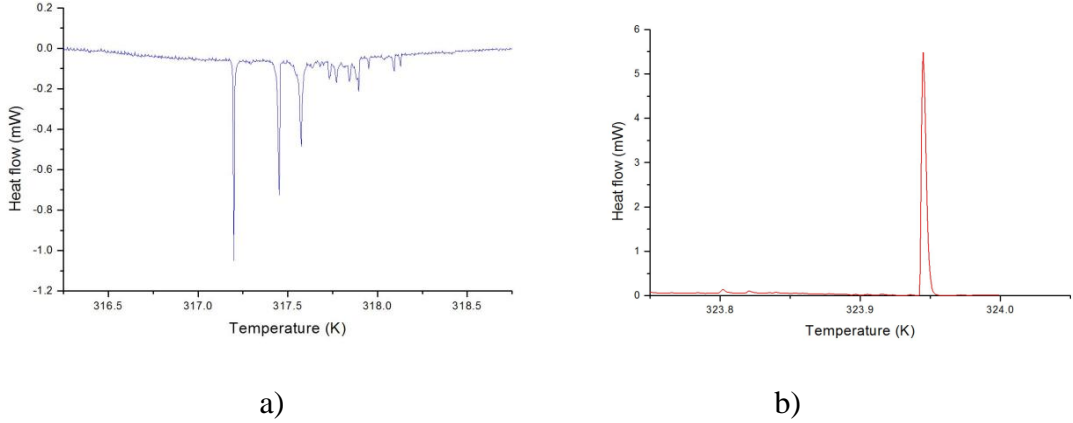

a)                                         b)

Fig.1. DSC results for cooling (a) and heating (b): heat flow versus temperature with 0.02 K/min rate on $Ni_2MnGa$ single crystalline (surface roughened) sample of 35 mg [2]. There is a significant difference between the two runs.

Thus it is not surprising that the above asymmetry is also manifested in differences of the acoustic and magnetic emission noise activities or numbers of hits as well as in the critical power law exponents of the probability densities of the peak energy and amplitude [1-4, 6-12], for cooling and heating. Fig. 3 illustrates this for the magnetic and acoustic activity during heating and cooling with 0.06K/min rate during the austenite/martensite, A/M, transformation in ferromagnetic $Ni_2MnGa$ single crystal [8]. Furthermore, Fig. 4 shows the energy distribution of acoustic emission, AE, in NiFeGaCo single crystals (the energy of an individual acoustic event, $E_j$, was determined from an approximate integration of the square of the AE voltage by its duration time [7]) for heating and cooling. More interestingly, according to an empirical observation [3], the asymmetries can be classified into two groups: *i)* the number of hits and the whole energy of acoustic emission is larger while the energy and amplitude exponents are smaller for cooling (positive asymmetry), *ii)* the situation is just the reverse.

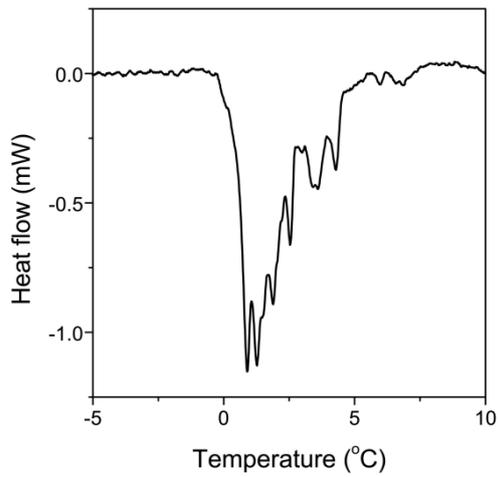
a)

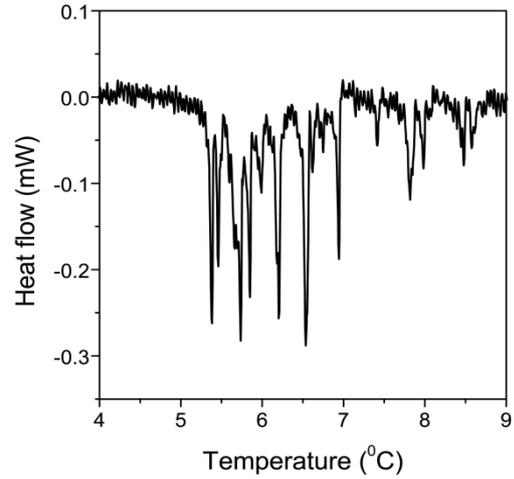
b)

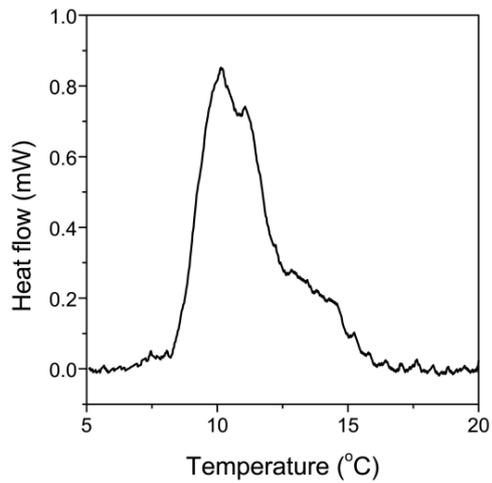
c)

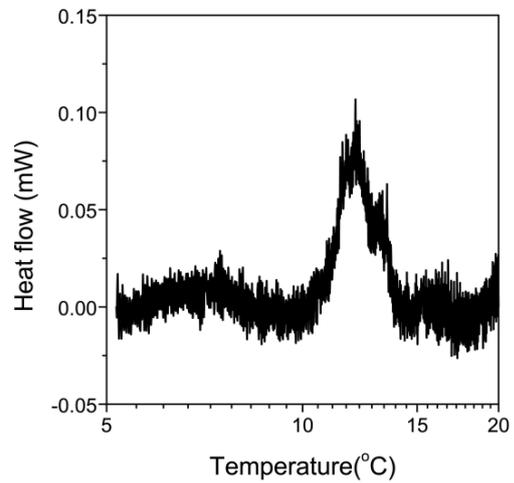
d)

Fig.2. DSC peaks obtained on NiFeGaCo single crystal No. 3: a) cooling with 3K/min, b) cooling with 0.3 K/min, c) heating with 3 K/min, d) heating with 0.3 K/min [3].

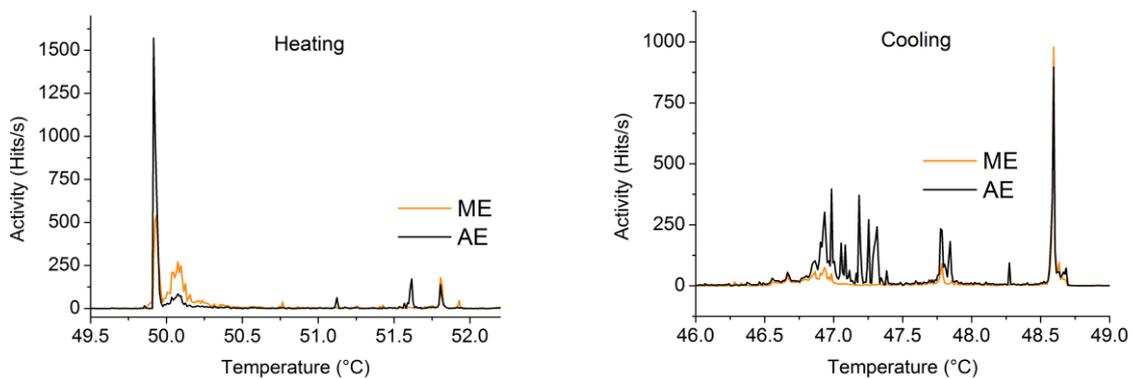

Fig.3. Magnetic and acoustic activity during heating and cooling, with 0.06 K/min rate, during the A/M transformation in $Ni_2MnGa$ single crystal [8].

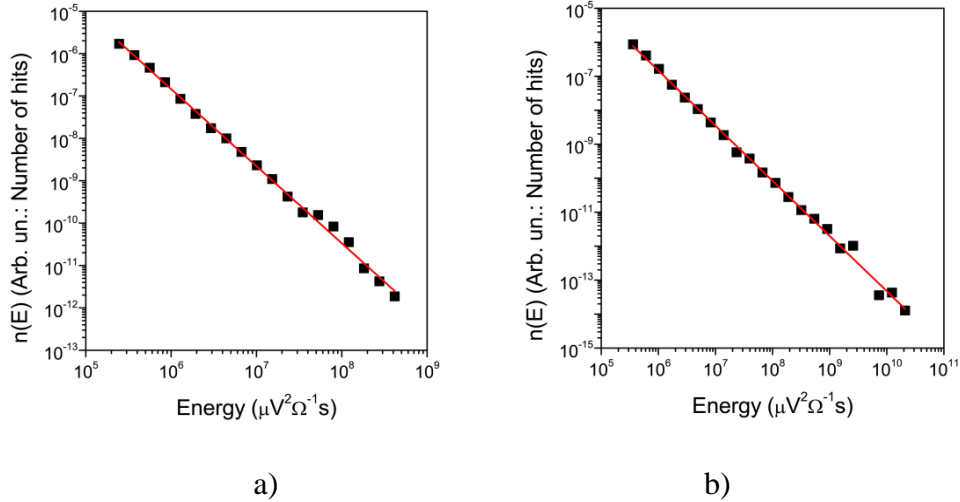

a)            b)

Fig.4. Acoustic energy distributions for NiFeGaCo single crystals for sample No 1. in [3]. The slopes of these straight lines are different for heating (a) and cooling (b): $\varepsilon_h=1.7$ and $\varepsilon_c=2.0$, respectively (see also Table I below).

      The origin of the above asymmetry, and especially the existence of the two different groups, is not clear yet, although its understanding would be very important to make the classification of the critical exponents proposed in [13] (see also [1]) unambiguous. The expected energy and amplitude exponents for different martensite symmetries should vary between $\varepsilon=2.0$-$1.6$, $\alpha=3.0$-$2.0$ from monoclinic to tetragonal symmetry (having intermediate value $\alpha=2.4$ for orthorombic martensite) [1,13]. On the other hand, as it was also pointed out in [2,3,7,8], the typical deviations between the corresponding exponents for cooling and heating are in the same range, which are the predicted differences due to different martensite symmetries.

      In this paper we provide an attempt for the interpretation of the above asymmetries, by considering possible different ways of partial relaxations of the elastic strain energy stored during the forward (cooling) and released during the reverse (heating) transformations. The elastic energy accumulates due to the transformation strain belonging to the A/M transformation: it is stored if this cannot be freely released (e.g. by forming surface steps) and the interaction/overlap of the elastic strain fields of growing martensite variants result in stored elastic strain energy. In the following the phrase "elastic energy" will be used in this sense. Our considerations will be primarily based on the analysis of the DSC, AE and magnetic emission, ME, results obtained in single crystalline samples. Note that, as it has been demonstrated recently [2,3,7,8,10,15,15], simultaneously measured DSC, acoustic as well as magnetic noise activities have quite a good coincidence with each other, confirming expectations that the thermal, acoustic and magnetic spikes have the same physical origin and related to the jerky character of interface motions during austenite/martensite transformations. Thus it is expected that the characteristics of the above asymmetry should have similar features for thermal, acoustic and magnetic noises.

      There is one very recent paper [17], dealing with the asymmetry of AE activity during forward and reverse transformations in CuZn(13.7at%)Al(17.0at%) and FePd(31.2at%) single

crystals. The authors came to the conclusion that the asymmetry could be a consequence of the fact that while nucleation is required for the transformation from the austenite to martensite phase, the reverse transition occurs by fast shrinkage of martensitic domains. Thus they argue that the asymmetry originates from the differences of the nucleation processes. As will be discussed, our approach is different and we argue that the differences in the elastic energy relaxations are at least as important as the differences in the nucleation during forward and reverse transformations.

Furthermore, our trial goes beyond the usual approach, namely that the heat measurable by DSC during cooling (index *c*) and heating (index h) can be given as [18,19]

$$Q_c = \Delta U_c + E_c + D_c \qquad (2)$$

and

$$Q_h = - \Delta U_c + E_h + D_h \qquad (3)$$

respectively, with the assumptions that $E_c = -E_h = E$ and $D_c = D_h$. Here $E_c$ and $D_c$ are the stored elastic as well as dissipated energies ($E_c>0$, $D_c>0$) during cooling, while $E_h$ and $D_h$ are the relaxed elastic as well as dissipative energies ($E_h<0$, $D_c>0$) during heating. $\Delta U_c = \Delta U$ and $\Delta U$ is the chemical (potential) energy change of transformation ($\Delta U = - L$, L is the latent heat), and $\Delta U<0$. According to (2) and (3) $(Q_c+Q_h)/2 = D_c$ and $(Q_h-Q_c)/2 = -\Delta U - E$. However, the accumulation as well as the release of the elastic energy (related to the transformation strain) during the down and up processes are usually accompanied by a partial relaxation of the elastic energy (e.g. by acoustic emission). Since these relaxation processes can be different for cooling and heating we will not suppose here that $E_c=-E_h=E$ and $D_c=D_h$ (see also below and the Appendix).

## 2. Calculations

### 2.1. Notations

Let us introduce the following notations:

- $\varepsilon_i$ and $\alpha_i$ are the energy and amplitude exponents according to equation (1), *i=h* and *c* for heating and cooling,
- $N_i$ and $E_{iAE}$ are the number of hits and the total energy of acoustic emission. For example $E_{cAE}$ is defined as $E_{cAE}=\sum_j E_{cj}$, where $E_{cj}$ denotes the energies of the individual AE peaks for cooling,
- $\gamma_\varepsilon = \frac{(\varepsilon_h - \varepsilon_c)}{\varepsilon_c}$, $\gamma_\alpha = \frac{(\alpha_h - \alpha_c)}{\alpha_c}$, $\mu = \frac{N_h}{N_c}$, $\xi = \frac{E_{hAE}}{E_{cAE}}$.

As it was mentioned earlier, for positive asymmetry $\gamma_\varepsilon, \gamma_\alpha > 0$ and $\mu, \xi < 1$.

- $E_{rc}$ is the elastic energy relaxed by AE during cooling,
- $E_t$ is the total elastic energy (which would be stored/relaxed without its relaxation)

- $E_{rh}$ is the elastic energy relaxed by AE during heating.

### 2.2. Possible ways of partial relaxations of the elastic energy

Let us start from the picture depicted e.g. in [20] (see also the Appendix): the storage/release of the elastic energy during the forward and reverse transformation itself is not an irreversible process, whereas the presence of local free energy barriers leads to irreversibility and intermittent dynamics (e.g. noises). Consider that during cooling and heating two types of acoustic sources (local free energy barriers) are operative (see e.g. [12] and [16,20]): i) usual frictional interactions of the moving interface (nucleation, pinning-depinning events): these are active in both directions, ii) during cooling or heating partial relaxations of the stored elastic energy (due to interaction/competitive growth of different martensitic variants) can occur in form of acoustic emissions, AE. Regarding the relative roles of these sources, while in [12] only the frictional interactions were mentioned, in Ni$_2$MnGa single crystalline samples it was demonstrated [16] that AE events were originated from specific local microstructural changes. It was concluded in [16] that contributions to AE from classical nucleation events could be excluded and the majority of the energy relaxations originated from the variant-variant interactions and from the interaction of martesite variants with grain boundaries (jamming effect) and pinning/depinning effects played only minor role. On the other hand, if the transformation took place towards single variant martensite structure the pinning and depinning were identified as dominant mechanisms of generation of AE (see e.g. [11]).

On the basis of the above observations we first assume (on the grounds of similar pinning/depinning events in both directions and neglecting the possible differences in the nucleations) that the energy dissipated by usual frictional motion of the interfaces is the same in both directions:

$$D_{fc} = D_{fh} = D. \qquad (4)$$

On the other hand one can write for the total energy measured by acoustic emission:

$$E_{cAE} = \delta_E E_{rc} + \delta_D D \qquad (5a)$$

and

$$E_{hAE} = \delta_E E_{rh} + \delta_D D, \qquad (5b)$$

for cooling and heating, respectively. Here $\delta_i < 1$ ($i=E,D$) denote the detected fraction of acoustic energy emitted (obviously, because of the detection losses, it is less than unity and in principle can be different for frictional interactions and elastic energy relaxations).

From (5a) and (5b) we can write

$$\xi = \frac{E_{hAE}}{E_{cAE}} = \frac{E_{rh} + \delta D}{E_{rc} + \delta D} . \qquad \text{with } \delta = \delta_D/\delta_E. \qquad (6)$$

It is expected that $\delta$ should be in the order of unity. Furthermore, let us define the fractions of the elastic energy relaxed by AE during cooling and heating as

$$\beta_c = \frac{E_{rc}}{E_t} \tag{7}$$

and

$$\beta_h = \frac{E_{rh}}{E_t - E_{rc}} = \frac{E_{rh}}{E_t(1-\beta_c)}, \tag{8}$$

respectively. In the denominator of (8) $E_t$ -$E_{rc}$ appears because this difference is the actual elastic energy stored during cooling and part of it can be relaxed by AE during heating. The expression (6), with (7) and (8), can be rewritten

$$\xi = \frac{\beta_h(1-\beta_c)+\delta\frac{D}{E_t}}{\beta_c+\delta\frac{D}{E_t}}, \tag{9}$$

from which the $\xi=1$ condition can be given as

$$\beta_h = \frac{\beta_c}{1-\beta_c}. \tag{10}$$

In the expressions (6) (or (9) the $E_{rc}$ and $E_{rh}$ energies (or the $\beta_c= E_{rc}/E_t$ and $\beta_h(1-\beta_c)=E_{rh}/E_t$ ratios) can be different because the elastic interactions at the moving interface as well as the development/regression of different martensite variants, leading to overlap/disintegration of their elastic field, can be different for cooling and heating and different parts of the elastic energy can be relaxed by emission of elastic waves (acoustic emission). Thus $\xi$ can differ from unity.

On the basis of (6) we arrive at the conclusion: *If the relaxed fraction of the total elastic energy, $E_t$ (which would be stored without relaxation during cooling), $E_r$, is larger/smaller than the corresponding relaxed fraction during heating, $E_{rh}$, then the asymmetry is positive/negative ($\xi <1$ or $\xi> 1$). The same statement is valid for the total energy of the acoustic emission peaks.*

In principle, besides the determination of the $\gamma_\varepsilon$, $\gamma_\alpha$, $\mu$ and $\eta$ parameters, the estimation of the values of $\beta_c$ and $\beta_h$ fractions, using DSC data (the heats during cooling and heating and the entropy of transformation) and some additional assumptions, is also possible (see the Appendix).

Note, that there are indications in the literature (see e.g. [8,12,15,21]) that acoustic and magnetic emission activities can also be observed after the martensite finish temperature during cooling. This indicates possible stress relaxations inside the freshly formed martensite even during cooling [8,12]. This fact can have an influence on the analysis presented before. Indeed in this case the values of $E_{rc}$ and $E_{rh}$ can be different and depend on the rate of changing the temperature and on the time, which the sample spent in martensitic state before heated. Thus the $\xi$ ratio can also depend on this time.

We can summarize the main arguments of this section as follows. During the motion of the martensite/austenite interface the development and release of the martensite structure (even without martensite variant rearrangements during cooling) leads to AE. Positive asymmetry ($\xi$, $\mu$ <1) can be observed if during the formation of the martensite multivariant structure additional rearrangements of the newly formed variants takes place (leading to considerable additional acoustic activity) and a more or less stable martensite structure transforms back during heating: in this case $N_h<N_c$ and $E_{hAE}<E_{cAE}$. At the same time the frictional contributions are less important (see eqns. (6) and (9)). In case of transformation by single interface motion less elastic energy accumulation/release is expected (there is only a minor stress accumulation during cooling due to the easy formation of the surface step at the moving interface, especially if the transformation takes place towards single variant martensite structure) and it is expected that both $\mu$ and $\xi$ will be close to unity. Thus negative asymmetry should be accompanied with some deviations from the above main effects.

If the nucleation effects, as suggested in [17], are important then one has to drop the assumption (4) and instead of it suppose that $D_{fc} \neq D_{fh}$ and $D$ has to be replaced by $D_{fh}$ and $D_{fc}$ in the nominator and denominator of (6) and (9), respectively. In this case $\xi$ can also be different from unity if these terms are important and the first terms are neglected. However, taking into account the observation of [17]: "An interesting feature is the fact that while the forward transition on cooling occurs by nucleation and growth of martensite variants, due to thermoelasticity, the reverse transformation occurs by variant shrinking.", one can arrive at controversial conclusion. Indeed on the basis of the above statement it would be expected that the AE activity should be higher for cooling than heating, i.e. $\xi<1$. This is on contrast to the observation of [17], where the AE activity was larger during heating. Nevertheless, emphasizing the asymmetry, it was concluded in [17] that "…the AE activity curves of the forward transition look more jerky-like than those corresponding to the reverse transition" (see also the Section 2.3 below).

### 2.3. Correlations between the noise parameters

Let us consider the experimentally observed correlations between the $\gamma_a$, $\gamma_c$, $\mu$ and $\eta$ parameters. Denoting by $n_i(E)$ the number of peaks of energy $E$, we can write for the total number of hits measured by AE as

$$N_i = \int n(E)_i dE, \tag{11}$$

and using that $n_i(E) \sim E^{-\varepsilon_i}$, if the cutoff region can be neglected (see e.g. the energy density functions shown in Fig. 3),

$$N_i \sim \int E^{-\varepsilon_i} dE, \tag{12}$$

is also valid. Furthermore, for the total energy of acoustic emission we have

$$E_{iAE} = \int n_i(E)E dE \sim \int E^{1-\varepsilon_i} dE. \tag{13}$$

One has to take the integral between $E_{max}=\infty$ and $E_{min}=E_m$, which is the minimal value of $E$ on the $n(E)$ function (see Fig. 4, where the numbers of hits are shown on the vertical axis): this is a certain lower bound to the power law behaviour as discussed in [22]). Since e.g. $\int E^{-\varepsilon_i} dE = \frac{E^{1-\varepsilon_i}}{1-\varepsilon_i}$ if $\varepsilon_i \neq 1$ ($\varepsilon_i$ is typically between 1.5 and 2.5 see Table 1 below) we have

$$\ln\mu = \ln\frac{N_h}{N_c} \sim -(\varepsilon_h - \varepsilon_c)\ln E_{mh} + \ln\frac{1-\varepsilon_c}{1-\varepsilon_h} + (1-\varepsilon_c)\ln\frac{E_{mh}}{E_{mc}} \cong$$

$$\cong -(\varepsilon_h - \varepsilon_c)\ln E_{mh} + \ln\frac{1-\varepsilon_c}{1-\varepsilon_h} \qquad (14)$$

and

$$\ln\xi = \ln\frac{E_{hAE}}{E_{cAE}} \sim -(\varepsilon_h - \varepsilon_c)\ln E_{mh} + \ln\frac{2-\varepsilon_c}{2-\varepsilon_h} + (2-\varepsilon_c)\ln\frac{E_{mh}}{E_{mc}} \cong$$

$$\cong -(\varepsilon_h - \varepsilon_c)\ln E_{mh} + \ln\frac{2-\varepsilon_c}{2-\varepsilon_h}. \qquad (15)$$

Here the terms proportional to $\ln\frac{E_{mh}}{E_{mc}}$ are neglected in both (14) and (15) because the $\frac{E_{mh}}{E_{mc}}$ ratio is close to unity. Our results, using the maximum likelihood method to data measured in single crystalline Ni$_2$MnGa and NiFeGaCo samples [2,3], showed that the optimal $E_m$ values were indeed very close for cooling and heating (see Fig. 6 in [3]) ), indicating that the appropriate threshold values were almost the same. According to (14) and (15), for $\varepsilon_h > \varepsilon_c$ both $\mu$ and $\xi$ are less than unity: this is the case of positive asymmetry. Finally, there exists a well-known scaling relation between the α and ε exponents [1]:

$$(\alpha - 1) = z(\varepsilon - 1), \quad (\text{with } z \cong 2). \qquad (16)$$

Thus, $\alpha \sim \varepsilon$ and $\gamma_\varepsilon \sim \gamma_\alpha$: for negative asymmetry $\gamma_\varepsilon$ and $\gamma_\alpha$ are negative as well as $\mu$ and $\xi$ are larger than unity.

### 3. Results and discussion

Before making a detailed analysis of the available experimental data it is worth emphasizing that there is a very good coincidence of the AE, ME and DSC peaks [4,8,10,15,16] obtained during heating or cooling. Even it was shown [4,8] that the critical energy exponents determined from the distributions constructed on DSC as well as AE data were the same within the experimental errors. In a more recent paper [8] the exponents of energy and amplitude distributions of the ME and AE signals were also determined. Thus the $\gamma_\varepsilon$ and $\gamma_\alpha$ parameters will be gathered from all available data of the above three types of measurements. It should be noted that, when one compares Fig. 1 and the AE results shown in Table I for the same sample (in the 2$^{nd}$ row) the number of peaks for cooling is larger than for heating in accordance with $\mu=N_h/N_c=0.84$ obtained from AE. On the other hand, an apparent contradiction can be meet, if the DSC curves shown Fig. 2 and the AE results shown in the last row of Table I are compared: while the number of DSC peaks are larger for cooling than heating, $\mu \cong 1$. This can be connected to the problem of experimental resolution of DSC peaks at the relatively large rates, which were available in the experiments: the decay time of the DSC peaks was relatively large, about 6 s [7], and overlapping of small peaks could happen (see also our comments to Fig. 6 below).

Table I Parameters characterizing the asymmetry in $Ni_2MnGa$ [2] and NiFeGaCo [3] single crystals. Note that the $\xi$ values were not given in [2] and calculated here from the measured data, according to the definitions $\xi = \frac{E_{hAE}}{E_{cAE}}$ and $E_{cAE} = \sum_j E_{cj}$.

| System | $\gamma_\varepsilon$ | $\gamma_\alpha$ | $\xi$ | $E_{hAE}$ (arb. units) | $\mu = N_h/N_c$ | $N_h$ | $\varepsilon_c$ | $\varepsilon_h$ | $\alpha_c$ | $\alpha_h$ | Ref. |
|---|---|---|---|---|---|---|---|---|---|---|---|
| $Ni_2MnGa$, single crystal, smooth AE | +0.27 | +0.41 | 0.12 | - | 0.40 | 7166 | 1.5 ±0.1 | 1.9 ±0.1 | 2.02 ±0.04 | 2.85 ±0.04 | 2 |
| $Ni_2MnGa$, single crystal, roughened, AE | +0.20 | +0.33 | 0.33 | - | 0.84 | 21274 | 1.5 ±0.1 | 1.8 ±0.1 | 2.03 ±0.04 | 2.70 ±0.04 | 2 |
| $Ni_2MnGa$, single crystal, roughened, DSC for cooling | - | - | - | - | - | - | 1.7 ±0.2 | - | - | - | 2 |
| $Ni_2MnGa$, single crystal, smooth ME | +0.27 | +0.29 | - | - | - | - | 1.5 ±0.1 | 1.9 ±0.1 | 2.25 ±0.15 | 2.90 ±0.2 | 8 |
| NiFeGaCo single crystal, (No.1) smooth AE | -0.05 | -0.20 | 6.9 | - | 1.1 | 1679 | 1.9 ±0.1 | 1.8 ±0.1 | 3.0 ±0.1 | 2.4 ±0.1 | 7 |
| NiFeGaCo single crystal, (No.1) roughened, AE, | -0.15 | -0.14 | 6.4 | $1.6 \times 10^{11}$ | 2.7 | 29436 | 2.0 ±0.1 | 1.7 ±0.1 | 2.9 ±0.1 | 2.5 ±0.1 | 3, 7 |
| NiFeGaCo single crystal, (No.2) roughened, AE, | 0.17 | 0.13 | 0.5 | $5.7 \times 10^{10}$ | 0.9 | 5135 | 1.6 ±0.1 | 1.8 ±0.1 | 2.4 ±0.1 | 2.4 ±0.1 | 3 |
| NiFeGaCo single crystal, (No.3) roughened, AE, | ~0 | ~0 | ~1 | $3.1 \times 10^{10}$ | ~1 | 11400 | 1.9 ±0.1 | 1.9 ±0.1 | 2.8 ±0.1 | 2.8 ±0.1 | 3 |

Since the two sets of measurements, made recently on single crystalline $Ni_2MnGa$ [2,8] and NiFeGaCo [3,7] samples in our group, represent more complete investigations from the point of view of the heating/cooling asymmetry let us first consider these data.

### 3.1. Results on $Ni_2MnGa$ single crystals

In ref. [2] calorimetric and acoustic emission studies were carried out on $Ni_2MnGa$ single crystals, with 10M martensite structure, at low cooling and heating rates (0.1 K/min and below). It was illustrated that, besides the low cooling and heating rates, the mass and the surface roughness were also important parameters in optimizing the best signal/noise ratio. We summarize here only the results obtained on the "Not treated" sample (the other two samples had different preliminary treatments to produce different twin structures and different behavior in martensitic state) at 0.1 K/min driving rate as shown in the first two rows of Table I for both original (smooth surface) and surface roughened (made by electro-erosion) samples [2]. In addition to the AE results, shown in Table I, it was also shown that the energy exponents obtained from DSC and AE runs during cooling ($\varepsilon=1.7\pm0.2$ as well as $\varepsilon=1.5\pm0.1$, respectively) were the same within the experimental errors (see the third row in Table I), confirming the results of [4] obtained in polycrystalline $Cu_{67.64}Zn_{16.71}Al_{15.65}$ samples.

In accordance with the observation on the AE activity made in [21] on similar *Ni₂MnGa* samples, the number of hits was higher for cooling i.e. the asymmetry is positive here. In addition to the results of AE the critical exponents of energy and amplitude of ME are also included in the fourth row of Table I at zero external magnetic field [8].

Interestingly, surface roughening has only a minor effect on the asymmetry: although the number of hits and the values of *ξ* and *μ* for AE are larger for surface roughened samples the values of $\gamma_\varepsilon$ and $\gamma_\alpha$ are slightly smaller, and the deviation is smaller than the estimated errors due to the uncertainties of exponents.

In [21] the observed asymmetry in the acoustic activity was attributed to the relaxation of the martensite structure by twinning, which is in qualitative agreement with our treatment presented in Sec. 1.2. Furthermore, in [16] it was discussed that AE events belonging to the activity of martensite variants played the dominating role as compared to the contributions from pinning/depinning effects in $Ni_2MnGa$ samples. This indicates that the second terms in the nominator and denominator of eqn. (6) can be neglected. In our recent paper [8] it was shown that the AE activity as the function of the martensite volume fraction, η, was stronger at larger η values and was different for cooling and heating (Fig. 5). This activity peak for cooling can also be an indication of the relaxation of the elastic energy by martensite rearrangement during the transformation and can be connected to the observations (see also [12,15,21]), that the asymmetry is accompanied with magnetic and acoustic noise activities even after the martensite finish temperature during cooling of $Ni_2MnGa$ alloys. In accordance with this, the positive asymmetry (*ξ is* less than unity) indicates that $E_{rc}/E_t$ is larger than $E_{rh}/E_t$ (or $\beta_h < \frac{\beta_c}{1-\beta_c}$) in (6) (or in (9), respectively).

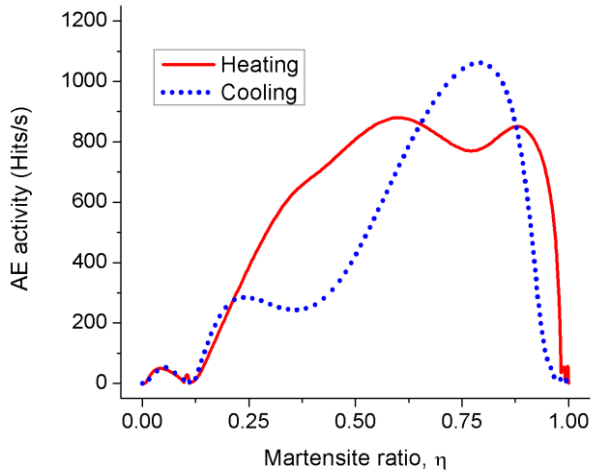

Fig.5. Acoustic activity versus the martensite volume fraction for cooling and heating in single crystalline $Ni_2MnGa$ [8].

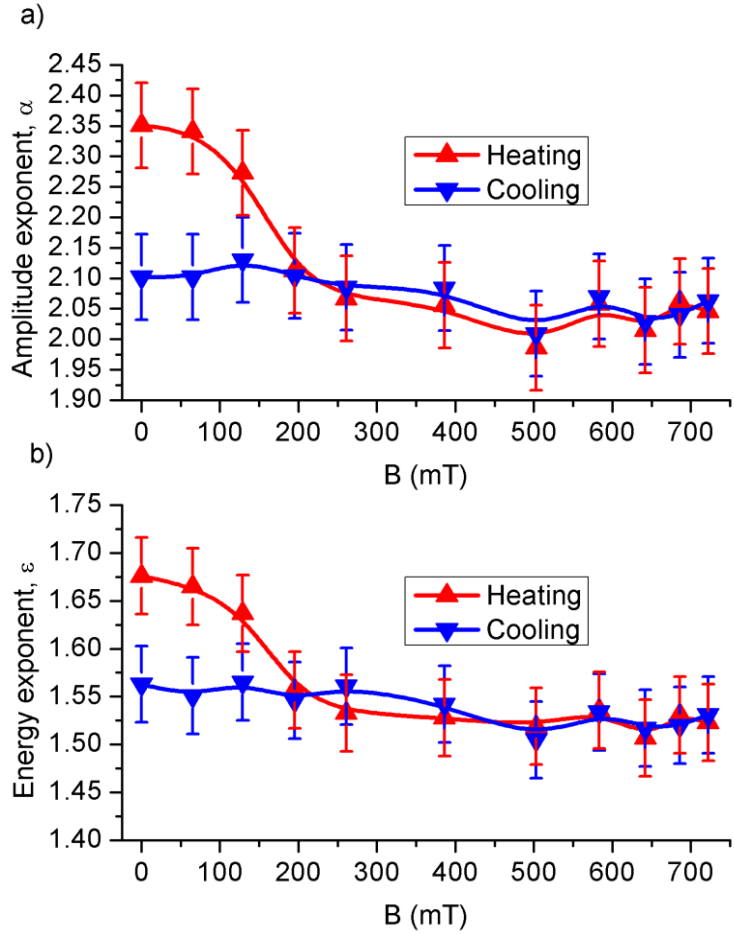

Fig. 6. Amplitude (a) and energy (b) exponents of AE as the function of the external magnetic field in single crystalline $Ni_2MnGa$ [8].

There is one more interesting result obtained in [8]. It is related to the effect of the constant external magnetic field. It was found that in both the AE and ME data the asymmetry observed at zero field disappeared with increasing magnetic field (Fig. 6) and this effect was attributed to the decreased multiplicity of the martensite variants. Indeed this transition was observed between 100 and 200 mT, which is in good agreement with the switching field (necessary to move the twin boundaries and start variant rearrangements) value obtained in [23] for the same samples. Thus, below this magnetic field values thermally induced multi-variant martensitic structure developed, while at higher field values a single variant structure (preferred by the magnetic field) developed and during this latter process less elastic energy accumulation is expected. This is in accordance with the disappearance of the asymmetry. However, interestingly only the critical exponents for heating showed changes and the exponents for cooling were unchanged (Fig. 6), although on the basis of the above arguments rather the cooling exponents should change. Understanding of this behaviour needs more detailed experiments providing more insight into the acoustic energy emission during micro/nano-structural changes of the development/regression of martensite variants during heating and cooling. While the positive asymmetry is observed at zero field (the elastic energy relaxation by AE is more pronounced during the development of the multi-variant martensite variant structure than during its regression), the question that why only the heating exponents decreased with increasing magnetic field is still not fully clear.

### 3.2. Results on NiFeGaCo single crystals

In [3] and [7] the effect of the presence of particles of $\gamma$-phase in NiFeGaCo single has been investigated on the mode of formation of the martensite phase: while in the homogeneous (free of precipitates) sample the transformation underwent by single interface motion (see Fig. 8 in [7]), in samples with precipitates many martensite needles were formed and grown in two specific directions (see Fig. 2 in [3]). Similarly the shape and the area of hysteresis curves, determined from DSC measurements were also different. These differences were also accompanied with differences in the asymmetries: in homogeneous crystals (Sample No.1, without $\gamma$-phase precipitates) positive, while in crystals with large $\gamma$-phase precipitates (5-15 μm, Sample No.2) negative asymmetry was observed. Results on aged crystals with bimodal structure (large and small $\gamma$-phase: particles 5-15 μm + 150-300 nm, Sample No. 3) were between these cases: the relative changes were practically zero. Table II summarizes details of the micro-structure of the three samples, while the last four rows of Table I show the characteristic asymmetry parameters. All the results (except sample 1, where a comparison of the AE results obtained on smooth and surface roughened samples were made: see the fifth and six rows in Table I) were obtained on surface roughened samples with 0.1 heating/cooling rates. The results nicely show that the micro structure and the mode of transformation (by single or multi interface motion) have definite influence on the type of asymmetry.

It can also be seen from Table I that on the sample 1 with smooth (polished) surface the asymmetry is a bit less, than for the surface roughened one. It illustrates that interestingly the introduction of more surface pinning/nucleation points, although the number events increased by more than one order of magnitude, does not considerably change the type of the asymmetry. This suggests that both types of AE sources (usual frictional effects and relaxations of the elastic energy) became more active and thus the characteristic asymmetry parameters remained almost unchanged.

Table II  Prehistory and structure of the investigated NiFeGaCo samples [7].

| Sample number | Preliminary heat treatment | Martensite crystal structure | Austenite crystal structure [4] | Size of the γ-phase particles |
|---|---|---|---|---|
| 1 | not treated | $L1_0$ tetragonal (or 14M monoclinic +$L1_o$) | $L2_1$ | - |
| 2 | 1373 K, 25 min | $L1_0$ tetragonal (or 14M monoclinic +$L1_o$) | B2 | 5-15 μm |
| 3 | 1373 K, 25 min, then 823 K, 30 min | $L1_0$ tetragonal (or 14M monoclinic +$L1_o$) | $L2_1$ | 5-15 μm + 150-300 nm |

Let us now consider the differences due to the presence of different precipitates. Fig. 7 illustrates that the splitting behaviour of the DSC curves as a function of heating rate is characteristically different for samples 2 and 3. While the microscopic images for these samples showed rather similar development of martensite needles, the bimodal microstructure in sample 3 results in much less split DSC peaks, indicating that the high number of nanosized pinning points probably causes many smaller elementary jumps (not resolvable in our device even at 0.1 K/min driving rate). This also should mean that the energy exponent and the number of hits of the AE noise for heating should be higher for sample 3 than for sample 2, due to the contribution of many jumps with small energy (see also Table I).

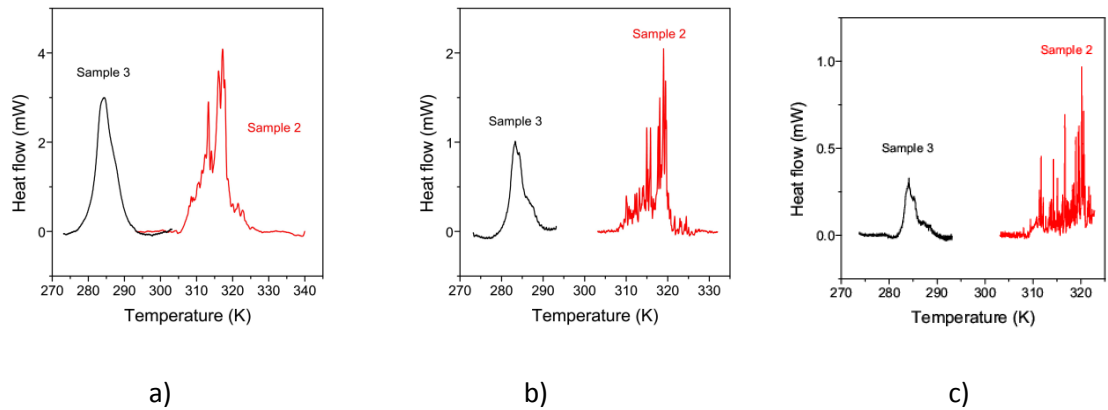

a) b) c)

Fig.7. DSC curves at different heating rates for the samples 2 and 3 [3]: a) 3 K/min, b) 1 K/min and c) 0.3 K/min [7].

The negative asymmetry in the homogeneous sample should be related to the single interface mode of the transformation, which suggests that during the motion of the single interface only a very moderate accumulation of the elastic energy, and thus its relaxation by AE, happens since during the motion the transformation strain can almost freely accommodated by the formation of the surface step accompanied with the single interface. Thus one would expect that the frictional term will be more important, leading to $\xi$ close to unity. As it can be seen in Table I this is not the case ($\xi=6.4$), which can be possible if the *relative change* of the very small elastic energy $E_t$ is high enough and larger for heating than cooling (see also the Appendix).

The positive asymmetry observed for sample 2 can have a similar interpretation than for the $Ni_2MnGa$ single crystal: the transformation mode is very similar (formation of many martensite needles) and in both cases the acoustic activity is larger for cooling as expected (see the explanation for $Ni_2MnGa$ too).

The zero asymmetry observed for sample 3 is an indication that increasing the number of nanosized pinning points causes many smaller elementary jumps of dissipation type, leading to determining role of the second terms in the nominator and denominator of (6).

It has to be noted that similarly to the effect of magnetic field on the asymmetry in $Ni_2MnGa$ single crystals, the full understanding of the reasons behind the appearance of three different behaviour calls for further, more detailed investigations.

### 3.3. Other literature data

Table III summarizes the experimental data available in the literature. It can be seen that indeed in a number of investigations the asymmetry was observed and only in AuCd alloys the asymmetry was negative ($\mu>1$). It is clear that more extended and detailed experimental data are desirable for arriving conclusions on the details of possible mechanisms behind. This is even so if consider the conclusions of the very recent paper [17] on the asymmetry in single crystalline CuZnAl and FePd samples. As it was already mentioned above the acoustic activity was different for heating and cooling and the $\xi$ values shown in Table 3 were calculated from the numbers of hits given in Table I of [17]. This value was 0.8 in FePd, and was practically independent of the heating/cooling rates (obtained at 0.1 K/min and 1K/min, respectively). On the other hand in CuZnAl the value of $\xi$ was sensitive to the threshold level: $\xi=1.7, 3.7$ and 4.3 for 30, 38, 40 dB thresholds, respectively (Table III contains the average of the values obtained at 38 and 40 dB). Furthermore in both crystals the energy and amplitude exponents were the same for cooling and heating, contradicting to the correlation between the asymmetry parameters predicted in Sec. 1.3 and observed in experiments on Ni$_2$MnGa and NiFeGaCo samples, as shown in Table II. It is worth mentioning that one of the most important conclusions on the difference between the behaviour of the CuZnAl and FePd samples in [17] was that while FePd displayed critical behaviour in both directions, deviation from the criticality was detected during heating in CuZnAl, although the authors could not discard the possible artefact due to the overlapping of small energy events.

Table III: Data collection on the observed asymmetry for the cooling/ heating process during martensitic transformations in different shape memory alloys (other than Ni$_2$MnGa and NiFeGaCo single crystals).

| System | $\gamma_\varepsilon$ | $\gamma_\alpha$ | $\xi$ | $\mu$ | ref. |
|---|---|---|---|---|---|
| Cu$_{67.64}$Zn$_{16.71}$Al$_{15.65}$ polycrystalline, AE | +0.05 | - | - | 0.27 | 4 |
| Cu$_{67.64}$Zn$_{16.71}$Al$_{15.65}$ polycrystalline, calorimetry | +0.05 | - | - | ~1 | 4 |
| Cu$_{69.3}$Zn$_{13.7}$Al$_{17.0}$ single crystal, AE | 0 | 0 | 4 | - | 17 |
| Cu-Al-Be, single crystal, Strain avalanches, mechanically induced transformation (the A→M transformation corresponds to cooling) | - | +0.33 | - | 0.75 | 9 |
| Ni$_{54.35}$Mn$_{23.18}$Ga$_{22.47}$ single crystal, ME | - | - | - | <1 | 10 |
| Fe-30%Pd single crystal, ME | - | - | - | >1 | 10 |
| Fe$_{68.8}$Pd$_{31.2}$ single crystal, AE | 0 | 0 | 0.8 | - | 17 |
| Fe$_{68.8}$Pd$_{31.2}$ single crystal, AE | 0.0 | 0.0 | - | ~1 | 11 |
| Fe$_{68.8}$Pd$_{31.2}$ polycrystal, AE | +0.26 | +0.38 | - | <1 | 11 |
| Au-47.5%Cd, polycrystal, AE | - | - | - | 2.9 | 12 |
| Au-47.5%Cd, single crystal, multiple interface, AE | - | - | - | 2.5 | 12 |
| Au-47.5%Cd, single crystal, single interface, AE | - | - | - | ~10 | 12 |

4. **Conclusions**

First quantitative attempt is offered for the interpretation of the asymmetry of forward and reverse martensitic transformations in shape memory alloys. It is based on energetic considerations, described in Sec. 1, and states that the asymmetry is positive ($\gamma_\varepsilon, \gamma_\alpha > 0$ and $\mu, \xi < 1$) if the relaxed fraction of the total elastic energy, $E_t$ (which would be stored without relaxation) during cooling, $E_{rc}$, is larger than the corresponding relaxed fraction during heating, $E_{rh}$. The same statement is valid for the total energy of the acoustic emission peaks (see eqns. (5) and (9)).

Comparison with experimental data in single crystalline $Ni_2MnGa$ and NiFeGaCo single crystals indicated that in most of the cases (except sample 3 for NiFeGaCo, where the presence of high number of nanosized pinning points caused many smaller elementary jumps of dissipation type) the contribution of frictional interactions of the moving interface (nucleation, pinning-depinning events) can be neglected and the differences in the relaxations of the elastic energy during cooling and heating play the determining role.

The effect of surface roughening, although it increased the number of events by more than one order of magnitude, does not considerably change the type of the asymmetry.

In the majority of samples investigated till now (Table I and III), the asymmetry is positive in accordance with the expectation that during cooling the elastic energy relaxations by AE are more considerable (due to the rearrangements of the newly formed martensite variants) than those during heating (when a more or less stable martensite structure transforms back).

The full understanding of the reasons behind the magnetic field dependence of asymmetry in $Ni_2MnGa$ single crystals as well as the appearance of the negative (or zero) asymmetry in NiFeGaCo single crystals with different microstructure calls for more detailed investigations.

**Acknowledgements**

The work was supported by the GINOP-2.3.2-15-2016-00041 project. The project was co-financed by the European Union and the European Regional Development Fund.

## Appendix
## Cooling

In general the storage/release of the elastic energy during the forward and reverse transformation itself is not an irreversible process, whereas the presence of local free energy barriers (related to friction on local external defects and to the relaxations of the elastic energy) leads to irreversibility and intermittent dynamics (e.g. noises) [20]. Let us consider the heat measurable in a DSC run. According to (3) we can write for the energy dissipated cooling as

$$D_c = D + E_{rc} = D + \beta_c E_t, \quad (A1)$$

(expressing that it contains two terms: the energy dissipated during the frictional motion of the interface and the fraction of the total elastic energy, which is relaxed by AE). Furthermore, since the elastic energy contribution to the DSC (the elastic energy stored during cooling) is given by

$$E_c = E_t - E_{rc} = E_t(1-\beta_c) \quad (A2)$$

eq. (2) has the form

$$Q_c = \Delta U_c + E_c + D_c = \Delta U + E_t + D \quad (A3)$$

and $\Delta U_c < 0$, $E_c$, $D_c > 0$). Eq. (A3) means that the heat measured by the DSC during cooling looks similar as there would not be any relaxation of the elastic energy during cooling (but $D$ differs from $D_c$).

## Heating

Now we should start from eqn. (3) and we can write

$$E_h = -(1-\beta_h)(E_t - E_r) = -(1-\beta_h)(1-\beta_c)E_t, \quad (A4)$$

since now the $\beta_h$ ($<1$) fraction of the stored elastic energy during cooling, $(E_t-E_r)$, is relaxed during heating in the form of AE ($E_h<0$). Furthermore

$$D_h = D + \beta_h(E_t - E_r). \quad (A5)$$

Thus

$$Q_h = -\Delta U - (1-\beta_h)(E_t - E_r) + D_f + \beta_h(E_t - E_r) = -\Delta U - E_t(1-\beta_c)(1-2\beta_h) + D. \quad (A6)$$

Furthermore

$$Q_h - Q_c = -2\Delta U - 2E_t + E_r + 2\beta_h(E_t - E_r) = -2\Delta U - 2E_t[1-\beta_h(1-\beta_c) - \beta_c/2] \quad (A7)$$

and

$$Q_h + Q_c = 2D + E_r + 2\beta_h(E_t - E_r) = 2D_f + E_t[\beta_c(1-2\beta_h) + 2\beta_h]. \quad (A8)$$

Now, $Q_c$, $Q_h$, $E_{hAE}$ as well as $E_{cAE}$ can be experimentally determined. In some cases, $\Delta U = T_o \Delta S$ ($T_o$ and $\Delta S$ are the equilibrium transformation temperature and the entropy of transformation) can also be obtained, but in these cases the hysteresis loops should have practically vertical branches [19, 24].

Thus in the analysis of experimental data relations (A3), A6) as well as (5a) and (5b) in the rewritten forms

$$E_{cAE} = \delta_E E_r + \delta_D D = \delta_E \beta_c E_t + \delta_D D \tag{A9}$$

$$E_{hAE} = \delta_E \beta_h (1-\beta_c) E_t + \delta_D D \qquad \delta_E, \delta_D, \beta_h, \beta_c < 1 \tag{A10}$$

can be used.

Carrying out simultaneous experiments by DSC and AE, the following parameters can be obtained: $Q_c$, $Q_h$, $E_{cAE}$, $E_{hAE}$, $\Delta S$, where $\Delta S$ is the entropy of transformation. Furthermore in special cases, as it was mentioned above, when the hysteresis curves have vertical branches [19,24,25] the equilibrium transformation temperature $T_o$ can also be estimated as $T_o = (M_s + A_f)/2$ ($M_s$ and $A_f$ are the martensite start and austenite finish temperatures) and thus $\Delta U = T_o \Delta S$ can also be obtained. But even in this case we have four equations in which there are six unknown parameters: $E_t$, $D$, $\delta_E$, $\delta_D$, $\beta_h$, $\beta_c$.

Table AI: Transformation temperatures, heats, entropies as calculated from DSC data [3]. Typical error of $Q_h$, $Q_c$ and $\Delta U$ is about ± 5-10 J/kg.

| Sample no. | $M_s$ (K) | $A_f$ (K) | $T_0$ (K) | $\Delta S$ (J/kgK) | $Q_h$ (J/kg) | $Q_c$ (J/kg) | $(Q_h+Q_c)/2$ (J/kg) | $(Q_h-Q_c)/2$ (J/kg) | $\Delta U$ (J/kg) |
|---|---|---|---|---|---|---|---|---|---|
| 1 | 280.5 | 286.5 | 283.5 | -12.3 | 3524 | -3436 | 44 | 3480 | -3487 |

Thus a detailed comparison with the experimental data cannot be carried out, unless we can find new relations or make further assumptions. The DSC results obtained in NiFeGaCo single [3] crystalline samples are summarized in Table AI. Since the hysteresis at heating rates 1 K/min in [3] had approximately vertical branches the estimation of the equilibrium temperature was possible from the $T_o \cong (M_s + A_f)/2$, relation [19, 24] the value of $\Delta U$ was also estimated. Let us try to make estimation on the $D/E_t$ ratio. Using the results shown in Table A1, from (A3) and (A6) we have:

$$E_t + D_f = 51 J/kg \tag{A11}$$

as well as

$$-E_t(1-\beta_c)(1-2\beta_h) + D_f = 37 J/kg. \tag{A12}$$

The difference of them leads to

$$E_t[1+(1-\beta_c)(1-2\beta_h)] = 14 J/kg. \tag{A13}$$

Since $\beta_c$, $\beta_h < 1$, this means that $14 J/kg > E_t > 7 J/kg$. Let us take $E_t \cong 10\ J/kg$. Then from (A11) we have $d = D/E_t \cong 4$. Now, assuming that the frictional term in $\xi$ is not important, i.e. we assume that the AE activity measured is dominantly due to stress relaxation effects related to the motion of surface steps and pinning effects result in much lower acoustic activity (see also Sec 2.2). Then (see also value of $\xi$ in the 6$^{th}$ row of Table I)

$$\xi = [\beta_h(1-\beta_c) + \delta D/E_t]/[\beta_c + \delta D/E_t] \cong \beta_h(1-\beta_c)/\beta_c = 6.4. \tag{A14}$$

Now, dividing (A13) by $E_t \cong 10\ J/kg$

$$0.4 = (1-\beta_c)(1-2\beta_h). \tag{A15}$$

From (A14) and (A15) we get $\beta_c = 0.04$ and $\beta_h = 0.29$, which - taking also into account that our results can be considered as only an order of magnitude estimates - are reasonable values.